\documentclass{article}

\usepackage[subpreambles=true]{standalone}
\usepackage{amsmath}

\usepackage{breqn}

\usepackage{csquotes}
\usepackage[backend=biber,sorting=none,style=trad-abbrv]{biblatex}
\usepackage{tabu}
\usepackage{tikz}

\numberwithin{equation}{section}

\AtEveryBibitem{
    \clearfield{note}
    \clearfield{pagetotal}
    \clearfield{isbn}
    \clearfield{url}
    \clearfield{doi}
    \clearfield{issn}
    \clearfield{urlyear}
    \clearfield{urlmonth}
    \clearfield{eprinttype}
    \clearfield{eprint}
    \clearfield{series}
    \clearfield{volume}
    \clearfield{place}
}

\begin{document}

\maketitle

\begin{abstract}

    In this paper we present a new, elementary derivation of non-relativistic spin using exclusively real algebraic methods. To do this, we formulate a novel method to decompose the domain of a real endomorphism according to its algebraic properties. We reveal non-commutative multipole tensors as the primary physically meaningful observables of spin, and indicate that spin is fundamentally geometric in nature. In so doing, we demonstrate that neither dynamics nor complex numbers are essential to the fundamental description of spin.

\end{abstract}

\section{Introduction}\label{sec:preliminaries}

The fundamental nature of spin is one of the most important topics in modern physics. It is most commonly viewed as a form of angular momentum intrinsic to particles and fields that aligns with a given axis in discrete amounts. This behaviour is widely considered an example of quantisation. In other settings, fundamental spins are the building blocks of emergent space-time, with deep connections to gravity as a result. However, studying spin in isolation from other areas of physics is challenging. This paper will present an elementary construction of all non-relativistic spins by real algebraic methods without the use of: complex numbers; manifolds; calculus; spinors; explicit matrix representations; quantum mechanical notions such as states or probabilities; or dynamical notions such as angular momentum, energy, or time. In so doing we will reveal a new perspective on its fundamental nature. To begin, let us examine the current mathematical formalism of spin.

\subsection{Current Mathematical Description of Spin}

The usual description of a spin-$s$ system is as a finite-dimensional, irreducible representation $\rho^{(s)}$ of the real Lie algebra $\sutwo$, of the Lie group $\specialunitarygroup{2}$. To find these representations, the traditional approach\cite{fulton_representation_1991} considers instead the algebra's complexification $\sutwoc$ and uses the ladder operator basis:
\begin{align*}
    \definition{\rho^{(s)}(\generator[\pm])&}{\sothreeladderoperatordefinitions[\rho^{(s)}]}\\
    \definition{\rho^{(s)}(H)&}{\rho^{(s)}(\generator[z])}
\end{align*}
\noindent to explore the representation's root system. 

While this method succeeds in finding every spin representation, we only gain limited insight into their fundamental nature. To see this, we note the usual interpretation of the ladder operators: they increase or decrease the amount of alignment or anti-alignment the spin angular momentum has with a particular spacial direction. This is a physically meaningful description, however the ladder operators are only definable in the complexification $\sutwoc$; as such their behaviour cannot form a foundational physical description of representations of the real $\sutwo$.

Another difficulty the standard formalism encounters is in revealing all the physically meaningful observables of the theory. To begin with, the ladder operators are not Hermitian, thus not observable. The two observables the formalism does highlight are, up to isomorphism, the generator $\rho^{(s)}(\generator[z])$ and the Casimir operator \eqref{eqn:casimir-phys}. Since in this picture these are sufficient to derive all of the spin representations, it is easy to believe that these are the only relevant ones. However, it is known that there are higher-order observables hidden in the spin matrices, such as the quadrupole and higher-order moments\cite{landau_quantum_1981}. Since these play no role in the traditional development of the spin theory, it is not clear if their existence is significant.

The physical significance of the group $\specialunitarygroup{2}$ is also unclear. Some insight can be gained by recognising that $\specialunitarygroup{2}$ is the double cover (and in this case also universal cover) of the homogeneous symmetry group of Euclidean three-space $\specialorthogonalgroup{3}$. Unlike $\specialunitarygroup{2}$, $\specialorthogonalgroup{3}$ has a direct physical interpretation as the group of rotations. Furthermore, $\sutwo\cong\sothree$ as Lie algebras. This connection between spin and geometric symmetry is not just mathematical, it is demonstrably non-trivial: fermionic systems require a $4\pi$ rotation to return to their original state.

From these observations, we state that a physically more meaningful description of spin will be derived by working exclusively with the natural structures associated with the rotation group $\specialorthogonalgroup{3}$, thus notions of dynamics shall be avoided. It is worth noting here that the rotations of $\specialorthogonalgroup{3}$ are not gradual transformations of a space over time, but atemporal mappings between two states of the space. As such they are adynamical, and their preservation of the Euclidean metric is what imparts their geometric character. As a symmetry of real three-space we will maintain this close link by avoiding the use of complex numbers.

This presents an immediate challenge, as without the algebraic closure of the complex numbers we have no guarantee of eigenvalues for our operators. This makes the usual root system analysis inaccessible. To overcome this difficulty, we will utilise entirely real algebraic methods, which we will show give a more meaningful description of spin entirely in terms of its physical observables. To proceed in this direction, let us first consider algebraic theories in physics more generally.

\subsection{Algebraic Theories in Physics}

The working definition of an algebraic theory we shall use throughout this work is as follows: an algebraic theory is an algebra over the field $\genericspacefield$ which completely describes the properties of the system of study within its structure. In particular, this means that all of the objects of the algebraic theory have physical significance, and that all of the states of the system can be described in terms of elements of the algebra. The first of these two points will be the focus of this paper, as each deserves careful discussion.

The pursuit of algebraic theories in physics was advocated by Einstein\cite{einstein_meaning_2013} as a means to more naturally describe quanta than a continuum theory. Though rarely used exclusively, many aspects of algebraic approaches have been behind several seminal results in physics. For example, Dirac's standard bra/ket\cite{dirac_principles_1981} enabled him to construct quantum theory almost entirely in terms of operators. Furthermore, his derivation of the Dirac equation from the Klein-Gordon equation necessitated the definition of an algebra between the $\alpha_{k}$ and $\beta$. The emergence of spin in that setting was not a result of relativity however, as \levyleblond\space derived the Pauli equation from the \schrodinger\space equation\cite{wilkes_pauli_2020} by similar means. This, again, required the acceptance of an algebraic structure inherent to the system he was describing. As a final historical example, Von Neumann defined an idempotent element with respect to the position-momentum algebra in his proof of the Stone-Von Neumann theorem\cite{v_neumann_eindeutigkeit_1931}. More recently, Hestenes\cite{hestenes_quantum_2020}, Doran and Lasenby\cite{doran_geometric_2003}, and Hiley and Callaghan\cite{hiley_clifford_2010,hiley_clifford_2010-1} have used Clifford algebras to study the \schrodinger, Pauli, and Dirac equations, indicating that more extensive algebraic study of quantum theory is possible.

In the case of non-relativistic spin, there already exist theories with strong algebraic influences such as Racah's spherical tensor operator formalism\cite{thompson_angular_2008}. However, these remain strongly bound to quantum mechanics and to its Hilbert space formalism, precluding the elementary study of spin that we are advocating. For spin-$\half$ and spin-$1$ though, there exist real algebraic constructions independent of quantum mechanics. These are the Clifford\cite{doran_geometric_2003}:
\begin{equation}\label{eqn:clifford}
    \sothreespinhalfalgebraidentityphys{a}{b}[\rho^{(\half)}]
\end{equation}
and Kemmer\cite{helmstetter_about_2010,micali_meson_2008}:
\begin{dmath}[center]\label{eqn:kemmer}
    \sothreespinonealgebraidentityphys{a}{b}{c}[\rho^{(1)}]
\end{dmath}
\noindent algebras respectively, where $\rho^{(s)}(S_{a}),\,a\in\set{1,2,3}$ are the usual matrix generators for spin-$s$\cite{binney_physics_2013}.

Together with the Lie bracket:
\begin{equation}\label{eqn:lie-bracket-rep-phys}
    \sothreeliebracketphys{a}{b}{c}[\rho^{(s)}].
\end{equation}
\noindent and the Casimir element:
\begin{equation}\label{eqn:casimir-phys}
    \sothreecasimirelementdefinition{a}[\rho^{(s)}]=s(s+1)I_{2s+1}
\end{equation}
\noindent these identities completely specify the properties of spin-$\half$ and spin-$1$, including the eigenspectrum of the generators $\rho^{(s)}(\generator[a])$. The algebras become completely real if we perform the transformation $\mapdefinition{\rho^{(s)}(\generator[a])}{\unitimaginary\rho^{(s)}(\generator[a])}$ as then \eqref{eqn:clifford}, \eqref{eqn:kemmer}, \eqref{eqn:lie-bracket-rep-phys}, and \eqref{eqn:casimir-phys} have all real constants. This demonstrates that real algebraic descriptions are possible for spin-$\half$ and spin-$1$. However, it is unclear how to generalise this description to arbitrary spin, or what the physical observables are in the theory.

To make progress, let us consider in the broadest terms what properties the spin representations share. All spin representations represent their generators by finite-dimensional matrices $\set{\rho^{(s)}(\generator[a])}$, or equivalently by their actions on spinors. The identity $I_{2s+1}$, and all linear combinations of matrix products of the $\set{I_{2s+1},\rho^{(s)}(\generator[a])}$ are also allowable actions on spinors. Finally, \eqref{eqn:lie-bracket-rep-phys} always holds. Implicit in this account are the following properties: the actions form a finite-dimensional vector space spanned by a subset of $\set{I_{2s+1},\rho^{(s)}(\generator[a]),\composition{\rho^{(s)}(\generator[a])}{\rho^{(s)}(\generator[b])},...}$; there exists an associative, bilinear product between them that respects \eqref{eqn:lie-bracket-rep-phys}; this vector space formed by the actions is closed under this product (by construction and finite-dimensionality\cite{axler_linear_2014}). More succinctly, the actions form a real unital associative algebra, where the commutator between two generators yields their Lie bracket. We shall use these observations as a heuristic to guide our arguments.

\subsection{Initial Setup for Algebraic Spin Analysis}

Let us now begin our analysis, starting from the Lie group of rotational symmetry $\specialorthogonalgroup{3}$. It is well known that $\specialorthogonalgroup{3}$ is generated by a real Lie algebra $\sothree$. $\sothree$ is a three-dimensional real vector space equipped with an alternating bilinear Jacobi product often called a Lie bracket. We will use the terms Lie bracket and Lie product interchangeably in this work. The Lie product of $\sothree$ is isomorphic to the cross-product; we adopt this notation to avoid confusion with commutators. In terms of the standard basis $\set{\generator[a]}$, the Lie product is:
\begin{equation}\label{eqn:lie-product}
    \sothreelieproductdefinition{a}{b}{c}.
\end{equation}
\noindent Note immediately that we are describing the real Lie algebra $\sothree$ and not its complexification $\sothreec$, which was described in \eqref{eqn:lie-bracket-rep-phys}; to return to the standard basis of $\sothreec$ simply transform:
\begin{equation}\label{eqn:physics-maths-conversion}
    \mapdefinition{\generator[a]}{-\unitimaginary\generator[a]}.
\end{equation}
\noindent All results will be derived using \eqref{eqn:lie-product}, so care should be taken to convert when needed.

Since we wish to describe spin with a unital associative algebra, let us first consider the most general unital associative algebra of the elements of $\sothree$. This is the tensor algebra\cite{bourbaki_algebra_1998} $\tsothree$ of $\sothree$:
\begin{equation}\label{eqn:tensor-algebra}
    \tsothree\cong\reals\oplus\sothree\oplus\tensorpower{\sothree}{2}\oplus\tensorpower{\sothree}{3}\oplus\dots.
\end{equation}
\noindent where $\tensorproduct$ is associative, bilinear, and has identity element $1$. The elements of this algebra are called \enquote{tensors}, and are all $\reals$-linear combinations of \enquote{$k$-adic} tensors:
\begin{equation*}
\begin{cases}
    \alpha\in\reals & k=0\\
    \displaystyle\bigotimes_{j=1}^{n}v_{j}=\tensor{v_{1};\dots;v_{n}},\,v_{j}\in\sothree & k\in\integers^{+}.
\end{cases}
\end{equation*}
\noindent We define the \enquote{tensor order} of a $k$-adic to be $k$, and extend to arbitrary linear combinations by taking the largest tensor order amongst the terms.

$\tsothree$ combines most of the properties common to spin representations except finite-dimensionality and encoding the identity \eqref{eqn:lie-product} within its commutator. It is at least clear how we may implement any one of these two properties. We may find a finite-dimensional algebra from $\tsothree$ by quotienting\cite{bourbaki_algebra_1998} out all tensors above a certain tensor order $k$:
\begin{equation}\label{eqn:tensor-algebra-quotient}
    \frac{\tsothree}{\ideal{\tensorpower{\sothree}{(k+1)}}}
\end{equation}
\noindent where $\ideal{\tensorpower{\sothree}{(k+1)}}$ is the ideal generated\cite{bourbaki_algebra_1998} by the order-($k+1$) tensors. On the other hand, we may impose the identity \eqref{eqn:lie-product} by constructing the universal enveloping algebra\cite{humphreys_introduction_1972} $\usothree$ of $\sothree$:
\begin{equation}\label{eqn:universal-enveloping-algebra}
    \usothree\cong\frac{\tsothree}{\ideal{\sothreeueaidentity{a}{b}}}
\end{equation}
\noindent where the $\ideal{\sothreeueaidentity{a}{b}}$ is the ideal generated by elements of the form in its argument. This embeds the Lie product \eqref{eqn:lie-product} into the commutator of the algebra. By construction, $\usothree$ is the most general associative algebra of the elements of $\sothree$ with \eqref{eqn:lie-product} embedded in this way, and thus all other algebras sharing these properties must derive from it. Abusing notation slightly we denote the product on $\usothree$ by $\tensorproduct$ as well, but which $\tensorproduct$ is intended will be clear from context. Like $\tsothree$, an arbitrary element of $\usothree$ can be written as a linear combination of $k$-adic tensors.

However, we derive mostly trivial algebras if we try to implement both finite-dimensionality and the identity \eqref{eqn:lie-product} by performing both quotients; more precisely this happens when the quotiented tensors are of order $2$ or greater. This is because in $\usothree$ the summands of \eqref{eqn:tensor-algebra} are no longer orthogonal. To see this consider that in $\usothree$:
\begin{equation}
    \tensorpower{\sothree}{2}\ni\tensor{a;b}-\tensor{b;a}=\sothreelieproduct{a}{b}\in\sothree.
\end{equation}
\noindent Informally the quotient fails because the ideal we construct in \eqref{eqn:tensor-algebra-quotient} does not respect the structure the Lie product identity gives the algebra.

The tools and procedures required to derive finite-dimensional algebras from $\usothree$ by quotient will be the focus of the remainder of this paper. In section \ref{sec:real-operator-formalism}, we will overcome the lack of eigenvalues by constructing a novel mathematical framework to decompose vector spaces according to the algebraic properties of real operators. In section \ref{sec:decomposition-of-universal-enveloping-algebra}, we will apply this formalism to $\usothree$ to re-express it as a direct sum of its physically meaningful components which respect the structure imparted by the Lie product. Finally, in section \ref{sec:spin-algebras} we will demonstrate how quotienting all but a finite number of these components leads to purely algebraic formulations of all spins, and necessarily a new physical description of it.

\section{Real Operator Formalism}\label{sec:real-operator-formalism}

We will now construct a novel, basis-independent, algebraic formalism to decompose a vector space according to the properties of a real operator on that space. Let $\genericspace$ be a vector space over $\genericspacefield$ and $A\in\genericspaceendomorphisms$. The minimal polynomial\cite{axler_linear_2014} of $A$ is a polynomial $n(x)$ of least order with coefficients in $\genericspacefield$ such that:
\begin{equation}\label{eqn:minpolproperty}
    n(A)=0_{\genericspace}
\end{equation}
\noindent on $\genericspace$, where $0_{\genericspace}$ is the zero map on $\genericspace$. It is unique up to scalar multiple and always exists when $\textup{dim}(V)\in\naturals$. Let us factorise the unique monic minimal polynomial $m(x)$ into a product of non-zero powers of irreducible polynomials over $\genericspacefield$:
\begin{equation}\label{eqn:factoredminpol}
    m(x)=\prod_{j=1}^{n}f_j^{d_j}(x)
\end{equation}
\noindent where $\forall j_{1}\neq j_{2},\,\,\gcd(f_{j_{1}}(x),f_{j_{2}}(x))=1$. As we are working with a monic polynomial, none of these factors are constant. Let us choose one value of $j=k$ and write:
\begin{subequations}
\begin{align}
    m(x)&=\pk{x}\qk{x}\label{eqn:minpolpq}\\
    \definition{\pk{x}&}{f_k^{d_k}(x)}\\
    \definition{\qk{x}&}{\smashoperator{\prod_{j=1,j\neq k}^{n}}f_j^{d_j}(x)}.
\end{align}
\end{subequations}
\noindent By \bezout's Identity\cite{tignol_galois_2001}, there exist polynomials $a_{k},b_{k}$ such that:
\begin{equation}
    \ak{x}\pk{x}+\bk{x}\qk{x}=\gcd(\pk{x},\qk{x})=1
\end{equation}
\noindent where $|a_{k}|+|p_k|<|m|$, $|b_{k}|+|q_k|<|m|$, and the final equality follows by construction. $a_{k}$ and $b_{k}$ may in general be computed by, for example, the extended GCD algorithm. We observe that the polynomial ring $\genericspacefield[A]$ is naturally ring isomorphic to a quotient ring of $\genericspacefield[x]$, since their polynomials differ only by $A\leftrightarrow x$ and the identity given by equation \eqref{eqn:minpolproperty}. This implies an identity in $\genericspacefield[A]$:
\begin{equation}
    \composition{\ak{A}}{\pk{A}}+\composition{\bk{A}}{\qk{A}}=\genericspaceidentitymap
\end{equation}
\noindent with $\circ$ denoting composition. For notational convenience let us define:
\begin{subequations}
\begin{align}
    \definition{\ik&}{\composition{\ak{A}}{\pk{A}}}\\
    \definition{\pik&}{\composition{\bk{A}}{\qk{A}}}\label{eqn:pikdefinition}.
\end{align}
\end{subequations}
\noindent By equations \eqref{eqn:minpolproperty} and \eqref{eqn:minpolpq} we see that $\ik$ and $\pik$ are orthogonal:
\begin{equation}\label{eqn:orthogonality}
    \composition{\ik}{\pik}=\composition{\pik}{\ik}=0_{\genericspace}.
\end{equation}
\noindent Additionally, equations \eqref{eqn:minpolproperty} and \eqref{eqn:orthogonality} imply idempotency:
\begin{subequations}
\begin{align}
    \composition{\pik}{\pik}&=\pik\\
    \composition{\ik}{\ik}&=\ik.
\end{align}
\end{subequations}
\noindent Thus, we have performed a partial orthogonal decomposition:
\begin{equation}
\begin{gathered}
    \genericspace\cong\image{\pik}\oplus\image{\ik}\\
    id_{\genericspace}=\pik+\ik
\end{gathered}
\end{equation}
where:
\begin{subequations}
\begin{align}
    \composition{\qk{A}}{\ik}&=0_{\genericspace}\\
    \composition{\pk{A}}{\pik}&=0_{\genericspace}
\end{align}
\end{subequations}
\noindent respectively. It can be proven that $\qk{x}$ is the minimal polynomial of $A$ on $\image{\ik}$; thus we may iterate this bipartite decomposition on $\image{\ik}$, until the final $q_{l}(x)$ has only a single multiplicand. This process is guaranteed to terminate since $|m|\in\integers^{+}$.

Thus, from the minimal polynomial of $A$ we have arrived at a basis independent orthogonal decomposition of $\genericspace$ through the projectors $\Pi_{j}$:
\begin{subequations}
\begin{gather}
    \genericspaceidentitymap = \sum_{j=1}^{n}\Pi_{j} \\
    \composition{f_{j}^{d_j}}{\Pi_{j}} = 0_{\genericspace}.
\end{gather}
\end{subequations}
\noindent with $\Pi_{j}$ defined as in equation \eqref{eqn:pikdefinition}. This has been achieved without the use of complex numbers, or reference to vectors in $\genericspace$; it has been derived entirely from the algebraic properties of $A$.

In the case where there is only one multiplicand in equation \eqref{eqn:factoredminpol} with $d_j>1$, we cannot find a resolution of the identity into more than one projector by this method alone. Furthermore, there may be subspaces closed under the action of $A$ within each $\image{\Pi_{j}}$ which we cannot differentiate using only the above. These limitations will not hamper our analysis of spin however, as we will find that every minimal polynomial encountered has $d_j=1$ for all factors. Furthermore, our method of decomposition will ensure that all subspaces closed under the action of $A$ are isolated.

It is instructive to relate what has just been developed to existing concepts. A traditional eigenspace is an $\image{\pik}:\composition{(A-\lambda\,\id)}{\pik}=0$, so $|f_{k}|=1$ and $|d_{k}|=1$. A generalised eigenspace\cite{axler_linear_2014} is an $\image{\pik}:\composition{(A-\lambda\,\id)^{d}}{\pik}=0$ where $d\in\integers^{+}$ is minimal, so $|f_{k}|=1$ and $|d_{k}|=d$. The case when $|f_{k}|>1$ does not occur when considering operators on complex vector spaces, as $\complexnumbers$ is algebraic closed. A simple example of a real operator with such a subspace is a planar rotation by an angle $\theta$: the subspace $\image{\pik}$ in the plane of rotation satisfies $\composition{(A^{2}-2\cos{\theta}A+\id)}{\pik}=0$, which is irreducible over $\reals$.

If instead of a single operator we have a finite collection of mutually commuting operators $\set{A_{n}}$, then each decomposition of the identity $\id_{\genericspace}=\sum_{a_{n}}\Pi_{a_{n}}(A_{n})$ we find using the above method may be composed together to give a unique mutual decomposition:
\begin{equation}
    \id_{\genericspace}=\smashoperator{\sum_{a_{1},a_{2},...}}\composition{\Pi_{a_{1}}(A_{1})}{\Pi_{a_{2}}(A_{2})}{...}
\end{equation}
\noindent Doing this, we cannot guarantee that the image of each combined projector is non-trivial; exactly which projectors have non-trivial image depends on the relationship between the operators in the collection. This is a basis-independent generalisation of simultaneous diagonalisation of operators to the real operator case, where even generalised eigenspaces may fail to exist. If the operators do not all mutually commute the situation is markedly more complex.

Of particular interest to our present problem is the case where all $|f_{k}|=1$ and $d_{k}=1$. In this case:
\begin{equation}
    \pk{x}=x-\lambda_{k}
\end{equation}
\noindent
and $|b_{k}|=0$, i.e. constant, so:
\begin{equation}
    \bk{A}=\frac{1}{\qk{\lambda_{k}}}
\end{equation}
The method of algebraic orthogonal decomposition of a vector space developed here, in the particular case of $|f_{k}|=1$ and $d_j=1$, will be used extensively in our analysis.

\section{Decomposition of the Universal Enveloping Algebra}\label{sec:decomposition-of-universal-enveloping-algebra}

We will now utilise the methods developed in section \ref{sec:real-operator-formalism} to decompose the universal enveloping algebra $\usothree$ into its irreducible, orthogonal components. Doing this will enable us to derive non-trivial, finite algebras from it by quotient.

\subsection{Actions on $\usothree$}

To begin the decomposition of $\usothree$ we must identify some suitable operators to use. The first natural Lie algebra action\cite{humphreys_introduction_1972} defined on $\usothree$ is left multiplication:
\begin{equation}\label{eqn:leftmultiplication}
    \definition{\sothreeleft[u]}{\mapdefinition{v}{\tensor{u;v}}}.
\end{equation}
\noindent Using left multiplication we can describe a recursive relationship between the $k$-adic and $(k-1)$-adic tensors:
\begin{subequations}
\begin{align}
    \tensor{v_{1};v_{2};...;v_{k}} &= \composition{\sothreeleft[v_{1}]}(\tensor{v_{2};...;v_{k}})\\
    v_{k}&=\sothreeleft[v_{k}](1).
\end{align}
\end{subequations}
\noindent We may use this relationship to aid our decomposition of an arbitrary $A\in\usothree$ in the following way. Since our method of decomposition is linear, we may decompose $A$ by decomposing its constituent $k$-adic tensors. The above recursive relationship shows that we may decompose a $k$-adic tensor by considering the action of left multiplication on the decomposition of a $(k-1)$-adic tensor. Therefore, to decompose $\usothree$ it is sufficient to start with a scalar and decompose the result of each left multiplication $\sothreeleft[v]$ by elements $v\in\sothree$. This allows us to build up our decomposition order-by-order.

Unfortunately, the left multiplication itself cannot also be used to decompose a finite-dimensional subspace of $\usothree$. This is because the tensor order of $\sothreeleft[v](A),\,\forall v\in\sothree,A\in\usothree$ is one greater than the tensor order of $A$; thus we have left the finite-dimensional subspace we were trying to study. If $\usothree$ were finite-dimensional this would not be an issue, as repeated left multiplications would eventually become linearly dependent. However, as $\usothree$ is infinite-dimensional no minimal polynomial for left multiplication by any non-zero element can be expected to exist.

To overcome this difficulty, we must utilise the adjoint action\cite{fulton_representation_1991}, the second natural Lie algebra action defined on $\usothree$:
\begin{equation}\label{eqn:adjointaction}
    \definition{\sothreead[u]}{\mapdefinition{v}{
        \begin{cases}
            uv & u\in\reals\\
            \tensor{u;v}-\tensor{v;u} & u\in\sothree\\
            \sothreead[a]\!\circ\!\sothreead[b] & u=\tensor{a;b}.
        \end{cases}
    }}
\end{equation}
\noindent The tensor order of $\sothreead[B](A),\,\forall A,B\in\usothree$ is the same as the tensor order of $A$, and thus we remain confined to the finite-dimensional subspace of study, provided that subspace is closed under the action of $\sothreead[B]$, which is the case for the $k$-adics. Specifically we will use the adjoint action of the centre of $\usothree$:
\begin{equation}
    \zusothree=\left\{z\in\usothree\,|\,\tensor{z;A}=\tensor{A;z},\,\forall A\in\usothree\right\}.
\end{equation}
\noindent The centre $\zusothree$ can be generated by sums and products of scalars $\reals$ and the Casimir element:
\begin{equation}
    \sothreecasimirelementdefinition{a}.
\end{equation}
\noindent The action of $\sothreead[\alpha],\forall\alpha\in\reals$ are just scalings, which act uniformly on all of $\usothree$. Thus, we need only consider the action of $\sothreead[\sothreecasimirelement]$. To this end, we introduce the following notation:
\begin{subequations}
\begin{gather}
    \definition{\casaction}{\sothreead[\sothreecasimirelement]}\\
    \definition{\casfactoraction{k}}{\sothreead[\sothreecasimirelement]+k(k+1)\,\usothreeid}.\label{eqn:cas-factor-action}
\end{gather}
\end{subequations}
\noindent The factor of $k(k+1)$ in \eqref{eqn:cas-factor-action} shall be explained shortly.

\subsection{Relationship Between $\sothreeleft[v]$ and $\casaction$}

As previously outlined, the key to decomposing $\usothree$ is to understand how the action of $\sothreeleft[v]$ for $v\in\sothree$ interacts with $\casaction$, as this enables us to decompose a $k$-adic tensor order-by-order starting from a scalar. This relationship is derived in Appendix \ref{app:proof-left-action-identity}:
\begin{equation}\label{eqn:leftactionidentity}
    \commutator{\casaction}{\commutator{\casaction}{\commutator{\casaction}{\sothreeleft[v]}}} + 2\commutator{\casaction^2}{\sothreeleft[v]}=0_{\usothree}
\end{equation}
\noindent where $\commutator{\cdot}{\cdot}$ is the commutator, and every implied product is composition. This identity holds on the whole of $\usothree$.

To understand the consequences of \eqref{eqn:leftactionidentity} it is instructive to consider its action on a subspace $\image{\Pi_{E+t}}$ for which $\composition{(\casaction+t\,\id)}{\Pi_{E+t}}=0$. Doing this, we find that a polynomial in $\casaction$ of the form:
\begin{equation}
    p(x)=(x+t)\big(x+(t+1+\sqrt{4t+1})\big)\big(x+(t+1-\sqrt{4t+1})\big)
\end{equation}
\noindent annihilates $\composition{\sothreeleft[v]}{\Pi_{E+t}}$. It can be proven that the two roots with radicals are natural numbers iff $t=m(m+1),\,m\in\naturals$. In that case:
\begin{equation}\label{eqn:general-decomposition-left-identity-polynomial}
    p(x)=(x+m(m+1))\big(x+(m+1)(m+2))\big)\big(x+(m-1)m)\big).
\end{equation}
\noindent For $m\neq0$ all three roots are consecutive naturals of the form $m(m+1)$, and when $m=0$ the roots are 0, 0, and 2.

To see the significance of this observation, let us begin our order-by-order decomposition by noting that for 0-adics $\alpha\in\reals$ and 1-adics $v\in\sothree$:
\begin{subequations}
\begin{align}
    \casaction(\alpha)&=0\\
    (\casaction+2\,\id)(v)&=0.
\end{align}        
\end{subequations}
\noindent
Therefore, on $\reals$ and $\sothree$ the action of $\casaction$ has minimal polynomials:
\begin{subequations}
\begin{align}
    m(x)&=x\label{eqn:minpolmonopole}\\
    m'(x)&=x+2\label{eqn:minpoldipole}
\end{align}
\end{subequations}
\noindent respectively. Since these contain a power of a single irreducible polynomial, no further decomposition can be made here. Noting that $0=0(0+1)$ and $2=1(1+1)$, we see that our iterative process of decomposition is initialised by subspaces annihilated by $\casaction+m(m+1)\,\id$ for some $m\in\naturals$. Thus, by \eqref{eqn:general-decomposition-left-identity-polynomial}, all non-trivial orthogonal subspaces in our decomposition are annihilated by compositions of $\casfactoraction{n}=\casaction+n(n+1)\,\id$ for various $n\in\naturals$. This accounts for the constant in \eqref{eqn:cas-factor-action}.

\subsection{Scheme of Decomposition}

Clearly from \eqref{eqn:minpolmonopole} and \eqref{eqn:minpoldipole}, minimality of an $\casaction$ polynomial of the form \eqref{eqn:general-decomposition-left-identity-polynomial} will depend on the subspace we are acting on. Regardless, we may still use it to gain some insight into the gross structure of $\usothree$: starting from $\reals$, let us alternately apply $\sothreeleft[v_{j}]$, then decompose the resulting subspaces according to \eqref{eqn:general-decomposition-left-identity-polynomial}. It is useful to capture each subspace derived this way as the image of a map from the $k$-order tensors $\tensorpower{\sothree}{k}\rightarrow\usothree$, where $k$ is the total number of $\sothreeleft[v_{j}]$ that have been applied. We note that for this to make sense, the domains of these maps are $\tensorpower{\sothree}{k}\subset\tsothree$.

On such a subspace $\image{\casfactorsubspace{k}}$, applying first $\sothreeleft[v]$ and then decomposing with $\casaction$ by the methods of section \ref{sec:real-operator-formalism}, we find $\forall B\in\tensorpower{\sothree}{m}$:
\begin{subequations}
\begin{equation}
    \composition{\sothreeleft[v]}{\casfactorsubspace{k}}(B) = \composition{\big(\stepdown{v}+\steplevel{v}+\stepup{v}\big)}{\casfactorsubspace{k}}(B)\label{eqn:identity-resolution-example}
\end{equation}
\begin{align}
    \definition{\composition{\stepdown{v}}{\casfactorsubspace{k}(B)}&}{
        \begin{dcases}
            0 & k=0 \\
            \composition{\frac{\composition{\casfactoraction{k}}{\casfactoraction{k+1}}}{4k(2k+1)}}{\sothreeleft[v]}{\casfactorsubspace{k}(B)} & k\in\integers^{+}
        \end{dcases}}\label{eqn:step-down-multipole}\\[0.5\baselineskip]
    \definition{\composition{\steplevel{v}}{\casfactorsubspace{k}(B)}&}{
        \begin{dcases}
            \composition{\frac{\composition{(\casaction-2\,\id)}{\casfactoraction{1}}}{-4}}{\sothreeleft[v]}{\casfactorsubspace{0}(B)} & k=0 \\
            \composition{\frac{\composition{\casfactoraction{k-1}}{\casfactoraction{k+1}}}{-4k(k+1)}}{\sothreeleft[v]}{\casfactorsubspace{k}(B)} & k\in\integers^{+}
        \end{dcases}
    }\label{eqn:step-level-multipole}\\[0.5\baselineskip]
    \definition{\composition{\stepup{v}}{\casfactorsubspace{k}(B)}&}{
        \begin{dcases}
            \composition{\frac{\composition{\casfactoraction{0}}{\casfactoraction{0}}}{4}}{\sothreeleft[v]}{\casfactorsubspace{0}(B)} & k=0\\
            \composition{\frac{\composition{\casfactoraction{k-1}}{\casfactoraction{k}}}{4(k+1)(2k+1)}}{\sothreeleft[v]}{\casfactorsubspace{k}(B)} & k\in\integers^{+}
        \end{dcases}
    }\label{eqn:step-up-multipole}
\end{align}
\end{subequations}
\noindent where we have introduced the \enquote{step-down/step-level/step-up by $v\in\sothree$} operators $\forall k\in\naturals$, which by construction satisfy:
\begin{subequations}
\begin{align}
    \composition{\casfactoraction{k-1}}{\stepdown{v}}{\casfactorsubspace{k}(B)}=0\\
    \begin{rcases}
        \composition{\casfactoraction{0}^2}{\steplevel{v}}{\casfactorsubspace{0}(B)} & k=0\\
        \composition{\casfactoraction{k}}{\steplevel{v}}{\casfactorsubspace{k}(B)} & k\in\integers^{+}
    \end{rcases}=0\\
    \composition{\casfactoraction{k+1}}{\stepup{v}}{\casfactorsubspace{k}(B)}=0
\end{align}
\end{subequations}
\noindent Note: if \eqref{eqn:general-decomposition-left-identity-polynomial} is not minimal on $\image{\composition{\sothreeleft[v]}{\casfactorsubspace{k}}}$, then some of these steps will be into the trivial subspace. This will prove to be the case for $\composition{\steplevel{v}}{\casfactorsubspace{0}}$.

\begin{figure}[H]
    \centering
    \makebox[0cm]{\begin{tikzcd}[column sep={1.9cm,between origins},row sep={1.9cm,between origins}, execute at end picture={
    \begin{scope}[on background layer]
        \coordinate (tlanchor) at ($(tl.south west)!0.5!(e0.north west)$);
        \node[rectangle, rounded corners, inner sep=0pt, outer sep=0pt, fill=green!10, fit=($(uso3.south east)!(tlanchor)!(uso3.north east)$)(m0)(uso3)] {};
        \coordinate (m01e) at ($(m0.south east)!0.5!(m1.north east)$);
        \coordinate (m01w) at ($(m0.south west)!0.5!(m1.north west)$);
        \node[rectangle, rounded corners, inner sep=0pt, outer sep=0pt, fill=yellow, nearly transparent, fit=(tlanchor)(e0)($(m0.south east)!0.5!(m1.north east)$)($(m01e)!(uso3.north east)!(m01w)$)(m0)] {};
        \coordinate (m12e) at ($(m1.south east)!0.5!(m2.north east)$);
        \coordinate (m12w) at ($(m1.south west)!0.5!(m2.north west)$);
        \node[rectangle, rounded corners, inner sep=0pt, outer sep=0pt, fill=yellow, nearly transparent, fit=($(e1.south west)!0.5!(e2.north west)$)(e2)($(m2.south east)!0.5!(m3.north east)$)($(m12e)!(uso3.north east)!(m12w)$)(m2)] {};
        \coordinate (m34e) at ($(m3.south east)!0.5!(m4.north east)$);
        \coordinate (m34w) at ($(m3.south west)!0.5!(m4.north west)$);
        \node[rectangle, rounded corners, inner sep=0pt, outer sep=0pt, fill=yellow, nearly transparent, fit=($(e3.south west)!0.5!(e4.north west)$)(e4)($(m4.south east)!0.5!(mdots.north east)$)($(m34e)!(uso3.north east)!(m34w)$)(m4)] {};
        \node[rectangle, rounded corners,  inner sep=0pt, outer sep=0pt, fill=yellow, nearly transparent, fit=($(bl.south east)!0.5!(reals.south west)$)(reals)($(areals.north east)!0.5!(aso3.north west)$)(areals)] {};
        \node[rectangle, rounded corners,  inner sep=0pt, outer sep=0pt, fill=yellow, nearly transparent, fit=($(so3.south east)!0.5!(so32.south west)$)(so32)($(aso32.north east)!0.5!(aso33.north west)$)(aso32)] {};
        \node[rectangle, rounded corners,  inner sep=0pt, outer sep=0pt, fill=yellow, nearly transparent, fit=($(so33.south east)!0.5!(so34.south west)$)(so34)($(aso34.north east)!0.5!(hdots.north west)$)(aso34)] {};
    \end{scope}
}]
    |[alias=tl]|&|[alias=areals]|&|[alias=aso3]|&|[alias=aso32]|&|[alias=aso33]|&|[alias=aso34]|&|[alias=hdots]|&&\\
    |[alias=e0]| \casfactoraction{0} & |[alias=im0]| \image{\multipole{0}} \arrow[rd] & - & V_{0} \arrow[rd] & W_{0} \arrow[rd] & X_{0} \arrow[rd, dotted] & ... \arrow[r, two heads, dashed, tail, "\cong"] & |[alias=m0]|\multipolecentralmodule{0}  \\
    |[alias=e1]| \casfactoraction{1} &   & \image{\multipole{1}} \arrow[ru, "\stepdown{v}"] \arrow[r, "\steplevel{v}"] \arrow[rd, "\stepup{v}"] & V_{1} \arrow[ru] \arrow[r] \arrow[rd] & W_{1} \arrow[ru] \arrow[r] \arrow[rd] \arrow[ru, shift left] \arrow[ru, shift right] \arrow[r, shift right] \arrow[r, shift left] \arrow[rd, shift right] \arrow[rd, shift left] & X_{1} \arrow[ru, dotted] \arrow[r, dotted] \arrow[rd, dotted] & ... \arrow[r, two heads, dashed, tail, "\cong"] & |[alias=m1]| \multipolecentralmodule{1}  \\
    |[alias=e2]|\casfactoraction{2} &   &   & \image{\multipole{2}} \arrow[ru] \arrow[r] \arrow[rd] & W_{2} \arrow[ru, shift right] \arrow[r] \arrow[rd] \arrow[ru] \arrow[r, shift right] \arrow[rd, shift right] & X_{2} \arrow[ru, dotted] \arrow[r, dotted] \arrow[rd, dotted] & ... \arrow[r, two heads, dashed, tail, "\cong"] & |[alias=m2]| \multipolecentralmodule{2}  \\
    |[alias=e3]|\casfactoraction{3} &   &   &   & \image{\multipole{3}} \arrow[rd] \arrow[r] \arrow[ru] & X_{3} \arrow[ru, dotted] \arrow[r, dotted] \arrow[rd, dotted] & ... \arrow[r, two heads, dashed, tail, "\cong"] & |[alias=m3]| \multipolecentralmodule{3}  \\
    |[alias=e4]|\casfactoraction{4} &   &   &   &   & \image{\multipole{4}} \arrow[ru, dotted] \arrow[r, dotted] \arrow[rd, dotted] & ... \arrow[r, two heads, dashed, tail, "\cong"] & |[alias=m4]| \multipolecentralmodule{4}  \\
    |[alias=edots]|\vdots &  &  &  &  &  & \ddots  & |[alias=mdots]|\vdots \arrow[d, "\oplus_{j=0}^{\infty}", Rightarrow, background color=green!10] \\
       |[alias=bl]|& |[alias=reals]|\reals & |[alias=so3]| \sothree & |[alias=so32]| \tensorpower{\sothree}{2} & |[alias=so33]| \tensorpower{\sothree}{3} & |[alias=so34]| \tensorpower{\sothree}{4} & ... \arrow[r,"\bigcup_{j=0}^{\infty}", Rightarrow, background color=green!10] & |[alias=uso3]| \usothree
\end{tikzcd}}
    \caption{Diagramatic representation of the decomposition of $\usothree$, where $\casfactoraction{j}\circ V_{j}=0$. The vertical yellow bands contain subspaces of a given tensor order. The horizontal yellow bands contain subspaces annihilated by a given polynomial of $\casaction$. The green vertical band contains the closures of the unions of all subspaces in each yellow band, expressed as modules $\multipolecentralmodule{k}$ of multipoles $\image{\multipole{k}}$ over the centre $\zusothree$. These objects will be defined in section \ref{sec:decomp-into-multipoles}.}\label{dia:step-decomposition}
\end{figure}
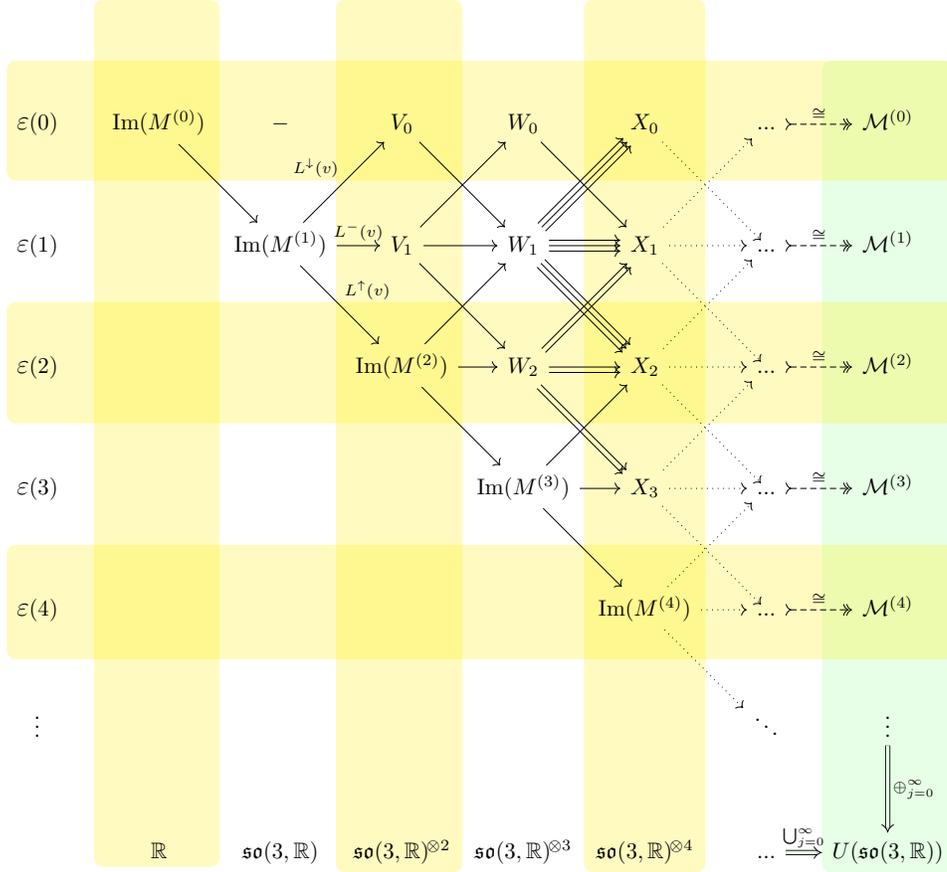

Using the step operators our decomposition scheme for $\usothree$ is equivalent to applying to 1 all sequences of steps by $v_{j}\in\sothree$ that yield non-trivial subspaces. We are guaranteed to decompose $\tensorpower{\sothree}{k}$ fully in terms of the non-trivial subspaces reached after $k$ steps due to \eqref{eqn:identity-resolution-example} holding at every step. This process is summarised graphically in Figure \ref{dia:step-decomposition}.

\subsection{Decomposition of $\usothree$ into Multipoles}\label{sec:decomp-into-multipoles}

From \eqref{eqn:general-decomposition-left-identity-polynomial}, \eqref{eqn:minpolmonopole}, and \eqref{eqn:minpoldipole}, we see $\forall k\in\naturals$ that amongst all subspaces reached in $k$ steps starting from 1, there is exactly one subspace annihilated by $\casfactoraction{k}$. This subspace was reached by stepping-up from the unique subspace reached in $k-1$ steps that is annihilated by $\casfactoraction{k-1}$. Furthermore, there are no subspaces annihilated by $\casfactoraction{k}$ reachable in fewer than $k$ steps.

This implies the existence of a family of maps:
\begin{equation*}
    \set{\mapdeclaration{\multipole{k}}{\tensorpower{\sothree}{k}}{\usothree}|\,k\in\naturals,\,\tensorpower{\sothree}{0}\cong\reals}
\end{equation*}
\begin{equation}
\begin{aligned}\label{eqn:recursivestepmultipole}
    \multipole{0}(\alpha)&=\alpha\\
    \multipole{k+1}(\tensor{v;B})&=\composition{\stepup{v}}{\multipole{k}(B)}
\end{aligned}
\end{equation}
\noindent whose images are the subspaces of least tensor order such that:
\begin{equation}
    \composition{\casfactoraction{k}}{\multipole{k}}=0.
\end{equation}
\noindent It can be proven that for each of these maps:
\begin{subequations}
\begin{gather}
    \forall A\in\usothree,\,\composition{\sothreead[A]}{\multipole{k}}=\composition{\multipole{k}}{\sothreead[A]}\label{eqn:multipoleadcommutation}\\
    \forall\tau\in S_k,\; \composition{\multipole{k}}{\tau}=\multipole{k}\label{eqn:multipoletotalsymmetry}\\
    \forall m\neq n\in \set{1,...,k},\;\sum_{a_m,a_n=1}^{3}\delta_{a_m a_n}\multipole{k}\Big(\bigotimes_{j=1}^{k}\generator[a_j]\Big)=0\label{eqn:multipolecontractionlessness}
\end{gather}
\end{subequations}
\noindent with \eqref{eqn:multipolecontractionlessness} recognised as vacuously true for $\multipole{0}$ and $\multipole{1}$. Given these properties, we find that \eqref{eqn:general-decomposition-left-identity-polynomial} is minimal on $\image{\composition{\sothreeleft[v]}{\multipole{k}}}$ for $k\in\integers^{+}$, implying that all $\multipole{k}$ have non-trivial image. We also find that \eqref{eqn:minpoldipole} is minimal on $\image{\composition{\sothreeleft[v]}{\multipole{0}}}$, implying:
\begin{subequations}
\begin{align}
    \composition{\steplevel{v}}{\multipole{0}} &=0\label{eqn:step-level-monopole}\\
    \composition{\stepup{v}}{\multipole{0}} &=\composition{\sothreeleft[v]}{\multipole{0}}.\label{eqn:step-up-monopole}
\end{align}
\end{subequations} 

Examining $\image{\multipole{k}}$ more closely, we find that it bears striking, but not exact, resemblance to the Cartesian $2^k$-pole tensor. For this reason we term the maps $\multipole{k}$ \enquote{multipoles}. As shown in Appendix \ref{app:multipole-images}, the images of the multipoles agree with the forms implied by \cite{coope_irreducible_1970}, though their algebraic properties and interrelationships are much clearer from this method.

The significance of the multipoles to our decomposition can be seen by considering the images of \eqref{eqn:step-down-multipole} and \eqref{eqn:step-level-multipole} on $\multipole{k}$, as derived in Appendix \ref{app:multipole-step-image-proof}. On $\multipole{0}$, $\stepdown{v}$ is trivial by definition and $\steplevel{v}$ is trivial by \eqref{eqn:step-level-monopole}. On $\multipole{1}$, the images are:
\begin{subequations}
\begin{align}
    \composition{\stepdown{\generator[a]}}{\multipole{1}(\generator[b])}&=\frac{1}{3}\sothreecasimirelement\delta_{ab}\label{eqn:step-down-dipole-image}\\
    \composition{\steplevel{\generator[a]}}{\multipole{1}(\generator[b])}&=\frac{1}{2}\sothreesum{c}\sothreestructureconstants{a}{b}{c}\generator[c]
\end{align}
\end{subequations}
\noindent and for $\multipole{k}$ with $k>1$ they are:
\begin{subequations}
\begin{equation}\label{eqn:step-down-multipole-image}
\begin{aligned}
    \composition{\stepdown{\generator[a]}&}{\multipole{k}\Big(\bigotimes_{j=1}^{k}\generator[b_{j}]\Big)}=\composition{\frac{L\big(4\sothreecasimirelement+(k-1)(k+1)\big)}{4(4k^{2}-1)}}{\sum_{p=1}^{k}\Bigg(\\ & (2k-1)\delta_{ab_{p}}\multipole{k-1}\Big(\bigotimes_{j\neq p}\generator[b_{j}]\Big)
    -\smashoperator{\sum_{q=1,q\neq p}^{k}}\delta_{b_{p}b_{q}}\multipole{k-1}\Big(\tensor{\generator[a];\smashoperator{\bigotimes_{j\neq p,q}}\generator[b_{j}]}\Big)\Bigg)}
\end{aligned}
\end{equation}
\begin{equation}\label{eqn:step-level-multipole-image}
    \composition{\steplevel{\generator[a]}}{\multipole{k}\Big(\bigotimes_{j=1}^{k}\generator[b_{j}]\Big)}=\half\sum_{p=1}^{k}\sothreesum{c}\sothreestructureconstants{a}{b_{p}}{c}\multipole{k}\Big(\tensor{\generator[c];\bigotimes_{j\neq p}\generator[b_{j}]}\Big).
\end{equation}
\end{subequations}
\noindent What this shows is that $\composition{\stepdown{v}}{\multipole{k}}$ and $
\composition{\steplevel{v}}{\multipole{k}}$ for $k>0$ can be written entirely in terms of $\multipole{k-1}$ and $\multipole{k}$ respectively. Since $\composition{\stepup{v}}{\multipole{k}}=\multipole{k+1}$ by definition, \eqref{eqn:identity-resolution-example} shows that $\composition{\sothreeleft[v]}{\multipole{k}(B)}$ can be written entirely in terms of multipoles $\forall k\in\naturals$.

Since in our decomposition we apply all combinations of steps starting from $\multipole{0}$, this means that every subspace we reach can be written entirely in terms of the multipoles. More precisely, we conclude that every subspace of $\usothree$ is a linear combination of products between the multipoles and central elements of $\zusothree$. This allows us to extend our proofs of minimality to all subspaces reached that are annihilated by the same $\casfactoraction{k}$ as $\multipole{k}$. In particular, this means that \eqref{eqn:step-level-monopole} and \eqref{eqn:step-up-monopole} hold on all subspaces annihilated by $\casfactoraction{0}$. This also allows us to linearly extend the step operators to arbitrary subspaces of $\usothree$. Being able to write all subspaces in terms of the multipoles is a manifestation of Weyl's theorem on complete reducibility\cite{hall_lie_2003}, since the multipoles are all simple as $\usothree$-modules under the adjoint action. 

With that, we have completed our decomposition of $\usothree$. In the process, we have identified a countable infinity of multipoles and given a recursive method for their construction. We have seen that all subspaces of $\usothree$ are isomorphic to direct sums of these multipoles, up to multiples of central elements $z\in\zusothree$. Our decomposition may thus be summarised:
\begin{equation}\label{eqn:multipoledecomposition}
    \usothree\cong\bigoplus_{j=0}^{\infty}\multipolecentralmodule{j}
\end{equation}
\noindent where $\multipolecentralmodule{j}=\big\{\sum_{p}\tensor{z_{p};m_{p}}\,|\,\forall z_{p}\in\zusothree,m_{p}\in\image{\multipole{j}}\big\}$.

\section{Spin Algebras}\label{sec:spin-algebras}

Unlike the tensor order decomposition \eqref{eqn:tensor-algebra} of $\usothree$, each summand in the multipole decomposition \eqref{eqn:multipoledecomposition} is orthogonal. Thus, we may derive a family of algebras:
\begin{equation}
    \definition{\spinalgebra{\half[k]}}{\frac{\usothree}{\ideal{\image{\multipole{k+1}}}}}.
\end{equation}
\noindent This process leaves only a finite number of basis elements, since \eqref{eqn:recursivestepmultipole} ensures that $\multipole{k+1}=0$ implies $\multipole{n}=0,\,\forall n> k+1$. What is not obvious is that this process will yield a real algebra, since each summand in \eqref{eqn:multipoledecomposition} is a module over $\zusothree$.

Appendix \ref{app:proof-scalar-casimir-action} proves that our quotient necessarily entails that:
\begin{equation}\label{eqn:spinalgebracasimiridentity}
    \sothreeleft[\sothreecasimirelement]=L\left(\frac{-k(k+2)}{4}\right)
\end{equation}
\noindent in the new algebra, i.e. the left action of the Casimir element $\sothreecasimirelement$ becomes the action of a real scalar. Reindexing $k=2s$, we see that the action of the Casimir element in algebra $\spinalgebra{s}$ is exactly what is expected for the spin-$s$ representation. In fact, this connection is total.

It can be proven that $\textup{dim}\left(\image{\multipole{k}}\right)=2k+1$, and thus $\spinalgebra{\half[k]}$ has dimension $\sum_{j=0}^{k}2j+1=(k+1)^2=(2s+1)^2$. This is exactly the complex dimension of the operators in the usual complexified spin-$s$ representation. Furthermore, $\image{\multipole{k+1}}=\set{0}$ implies that if $k$ is odd:
\begin{equation}\label{eqn:half-spin-eigenspectrum}
    \bigotimes_{j=0}^{\half (k-1)}\left(\tensor{\generator[a];\generator[a]}+\big(j+\half\big)^2\right)=0
\end{equation}
\noindent
and if $k$ is even:
\begin{equation}\label{eqn:whole-spin-eigenspectrum}
    \tensor{\generator[a];\bigotimes_{j=1}^{\half k}}\left(\tensor{\generator[a];\generator[a]}+j^2\right)=0
\end{equation}
\noindent
which yields the complete eigenspectrum expected for our basis. Due to this correspondence we will name the algebra $\spinalgebra{s}$ the \enquote{spin-$s$ algebra}. More concretely, the spin-$s$ representation is simply an associative algebra representation of $\spinalgebra{s}$, and derives its bulk structure from it.

In deriving the algebras $\spinalgebra{s}$, we have established that any spin may be specified entirely by its largest non-zero multipole, and equivalently described by: a finite collection of multipoles $\set{\multipole{n}|\,n\in\set{0,...,2s}}$; their multiplication table; and the implied relation $\leftcas=\sothreeleft[-s(s+1)]$. Such a multiplication table is given in Appendix \ref{app:multipole-multiplication}.

However, there are certain aspects of the usual formalism implied but not immediately accessible in the algebraic theory. For example, the non-zero parts of the eigenspectra from \eqref{eqn:half-spin-eigenspectrum} and \eqref{eqn:whole-spin-eigenspectrum} are pure imaginary; this means projectors into the corresponding eigenspaces are not constructable within the real $\spinalgebra{s}$. Similarly, the matrix representations of the odd $n$ multipoles are anti-Hermitian, while those of the even $n$ multipoles are Hermitian.

This is not a defect within the spin algebra formalism however; it indicates that the observable multipoles and spin eigenstates are the result of coupling with complex structure from another algebra within a larger physical theory. Therefore, the observability of these objects is an emergent, non-trivial prediction of such a theory.

For example, if we follow standard quantum mechanics and construct the algebra tensor product between $\spinalgebra{s}$:
\begin{equation}\label{eqn:spin-algebra-heisenberg}
    \spinalgebra{s}\boxtimes\heisenbergassociativealgebra=\Big\{\sum_{j}A_{j}\boxtimes B_{j}\,\big\vert\,(A_{j}\boxtimes B_{j})\boxtimes(A_{k}\boxtimes B_{k})=(\tensor{A_{j};A_{k}})\boxtimes(B_{j}\hat{\otimes}B_{k})\Big\}
\end{equation}
\noindent $\forall A_{j}\in\spinalgebra{s},B_{j}\in\heisenbergassociativealgebra$, and the associative Heisenberg algebra:
\begin{equation}
    \definition{\heisenbergassociativealgebra}{\frac{U(\heisenbergliealgebra)}{\ideal{\centralgenerator\hat{\otimes}\centralgenerator+\hbar^2}}}
\end{equation}
\noindent with $\hbar\neq0$, which is formed from the real Heisenberg lie algebra,
$\heisenbergliealgebra=\textup{span}\big(\heisenbergalgebragenerators\big)$:
\begin{equation}
\begin{gathered}
    \positionmomentumbracket{j}{k}\\
    \positioncentralbracket{j}\\
    \momentumcentralbracket{j}
\end{gathered}
\end{equation}
\noindent we find we are now able to factorise previously irreducible polynomials:
\begin{equation}
\begin{aligned}
    (\tensor{\generator[a];\generator[a]}+\alpha^2)\boxtimes1 &= (\tensor{\generator[a];\generator[a]})\boxtimes\Big(\frac{\centralgenerator\hat{\otimes}\centralgenerator}{-\hbar^2}\Big)+\alpha^2\boxtimes1\\
    &= -\frac{1}{\hbar^2}\big((\generator[a]\boxtimes\centralgenerator)\boxtimes(\generator[a]\boxtimes\centralgenerator)-(\alpha^{2}\hbar^2)1\boxtimes1\big)\\
    &= -\frac{1}{\hbar^2}\big(\generator[a]\boxtimes\centralgenerator-(\alpha\hbar)1\boxtimes1\big)\boxtimes\big(\generator[a]\boxtimes\centralgenerator+(\alpha\hbar)1\boxtimes1\big).
\end{aligned}
\end{equation}
\noindent From the identities \eqref{eqn:half-spin-eigenspectrum} and \eqref{eqn:whole-spin-eigenspectrum}, this implies that $\generator[a]\boxtimes\centralgenerator$ has the usual real eigenspectrum expected for spin operators in quantum physics:
\begin{equation}
    \set{-s\hbar,\,-(s-1)\hbar,\,\dots,\,(s-1)\hbar,\,s\hbar}
\end{equation}
\noindent and reveals why this eigenspectrum has units of $\hbar$. Furthermore, $\generator[a]\boxtimes\centralgenerator$ has Hermitian matrix representations. Together, these observations reveal that $\generator[a]\boxtimes\centralgenerator$ has similar properties to the angular momentum operators, whereas we have seen that $\generator[a]\boxtimes1$ does not. Therefore, we can say that the character spin has in quantum mechanics, as a form of angular momentum, is an emergent property of the algebra coupling \eqref{eqn:spin-algebra-heisenberg}; it is a non-trivial \textit{prediction} of quantum mechanics, not an intrinsic property of spin.

From the above, it also follows that, by their total symmetry, all multipoles formed from $\generator[a]\boxtimes\centralgenerator$ also have Hermitian matrix representations. The measurability of these observables in experiment will ultimately depend on the precise form of their electromagnetic couplings; however one might expect that the coupling strength of the $2^{k}$-pole for a particle of charge $e$ and mass $m$ would be of the order:
\begin{equation}
    \left(\frac{e\hbar}{2m}\right)^k
\end{equation}
\noindent to be consistent with the norms of the $(\generator[a]\boxtimes\centralgenerator)$-multipoles and the coupling strength of the spin magnetic dipole moment.

Independently of which algebras we might couple $\spinalgebra{s}$ to, we have established that it captures the essential structure of a spin-$s$ system. By using our real algebraic methods we have shown that this structure can be derived without the use of dynamics, matrix representations, or complex numbers, amongst other things. Therefore, by only using those structures naturally associated with the geometric symmetry group $\specialorthogonalgroup{3}$ of Euclidean three-space, we must conclude that spin is similarly geometric in nature.

\section{Conclusion}

In this paper we have constructed a completely algebraic theory of non-relativistic spin from spacial symmetry using only elementary arguments, without the use of: quantisation, dynamics, calculus, matrix representations, or complex numbers. To do this we developed a formalism appropriate to the study of real operators, which can readily be applied to study other symmetries, such as those arising in field theory, amongst other mathematical and physical contexts. Through this formalism, we have shown that a spin-$s$ system is a finite collection of non-commutative generalisations of Cartesian multipole tensors, and completely determined by specifying only the largest non-zero multipole. In working exclusively with structures naturally related to a geometric symmetry group, we have indicated that spin is fundamentally geometric in nature.

\section{Acknowledgements}

The author wishes to acknowledge B. J. Hiley, P. Van Reeth, A. Nico-Katz, G. Van Goffrier, and M. Hajtanian for their insightful discussions and comments.

\printbibliography
\appendix

\section{Table of Images of Multipoles}\label{app:multipole-images}

\begin{table}[H]
    \centering
    \makebox[0cm]{
        
\begin{math}
    \tabulinesep = 0.5\baselineskip
\begin{tabu}{c |[1pt] c}
    \text{Multipole}& \text{Image of $\tensor{\generator[a];\generator[b];...}$}\\\tabucline[1pt]{-}\everyrow{\tabucline{-}}
\multipole{0} & \sothreemonopoledefinition \\
\multipole{1} & \sothreedipoledefinition{a} \\
\multipole{2} & \displaystyle\sothreequadrupoledefinitioncompact{a}{b} \\
\multipole{3} & \displaystyle\sothreeoctupoledefinitioncompact{a}{b}{c} \\
\multipole{4} & \displaystyle\sothreehexadecapoledefinitioncompact{a}{b}{c}{d}\everyrow{}
\end{tabu}
\end{math}

    }
    \caption{Images of the multipoles $k=0,...4$ on $k$-adic tensors, where $S(A)$ is the set of permutations of the set $A$}
\end{table}

\section{Table of Multiplication for Multipoles}\label{app:multipole-multiplication}

\begin{table}[H]
    \centering
    \makebox[0cm]{
        \resizebox{1.7\linewidth}{!}{

\begin{math}
    \tabulinesep = 0.5\baselineskip
    \begin{tabu}{c |[1pt] c | c | c }
        \tensor{R;C} & \multipoletensor{} & \multipoletensor{e} & \multipoletensor{ef} \\\tabucline[1pt]{-}\everyrow{\tabucline{-}}
        \multipoletensor{} & \multipoletensor{} & \multipoletensor{e} & \multipoletensor{ef} \\
        \multipoletensor{a} & \multipoletensor{a} & \dipoledipole{a}{e}{s} & \dipolequadrupole{a}{e}{f}{s} \\
        \multipoletensor{ab} & \multipoletensor{ab} & \quadrupoledipole{a}{b}{e}{s} & \quadrupolequadrupole{a}{b}{e}{f}{s}
    \end{tabu}
\end{math}

        }
    }
    \caption{A partial table of multiplication for multipoles, where $\definition{\multipoletensor{a_{1}a_{2}...a_{k}}}{\multipole{k}(\tensor{\generator[a_{1}];\generator[a_{2}];...;\generator[a_{k}]})}$. Repeated indices in the same term are summed over. This may be extended using the images of multipoles in Appendix \ref{app:multipole-images}, and the results of Appendix \ref{app:multipole-step-image-proof}.}
\end{table}

\section{Proof of Left Action Identity}\label{app:proof-left-action-identity}

To facilitate this and other proofs we must first discuss some identities. Let us define the right multiplication:
\begin{equation}
    \sothreeright[v]=\mapdefinition{A}{\tensor{A;v}}
\end{equation}
\noindent where we note this is not a Lie algebra action on $\usothree$, since:
\begin{equation}
    \composition{\sothreeright[a]}{\sothreeright[b]}-\composition{\sothreeright[b]}{\sothreeright[a]}=\sothreeright[\sothreelieproduct{b}{a}]\neq\sothreeright[\sothreelieproduct{a}{b}].
\end{equation}
\noindent This allows us to describe the adjoint action of $v\in\sothree$:
\begin{equation}\label{eqn:adjoint-l-r}
    \sothreead[v]=\sothreeleft[v]-\sothreeright[v]
\end{equation}
\noindent Noting $\forall A,B\in\usothree$:
\begin{equation}
    \commutator{\sothreeleft[A]}{\sothreeright[B]}=0
\end{equation}
\noindent we easily see the commutators:
\begin{gather}
    \commutator{\adgen{a}}{\leftgen{b}}=\commutator{\leftgen{a}}{\leftgen{b}}=\leftliegen{a}{b}\label{eqn:adjoint-l-commutation}\\
    \commutator{\adgen{a}}{\rightgen{b}}=\commutator{-\rightgen{a}}{\rightgen{b}}=\rightliegen{a}{b}
\end{gather}
Next, let us examine $\casaction$ more closely.
For central elements $z\in\zusothree$:
\begin{gather}
    \commutator{\sothreeleft[z]}{A}=0\label{eqn:left-central}\\
    \sothreeleft[z]=\sothreeright[z].
\end{gather}
\noindent Then:
\begin{equation}\label{eqn:cas-action-identity}
    \casaction=\sothreesum{a}\composition{\adgen{a}}{\adgen{a}}=2\sothreeleft[\sothreecasimirelement]-2\sothreesum{a}\composition{\leftgen{a}}{\rightgen{a}}.
\end{equation}

We may now proceed with the proof.
\begin{align*}
    \commutator{\casaction}{\leftgen{b}} &= \sothreesum{a}\commutator{\composition{\adgen{a}}{\adgen{a}}}{\leftgen{b}}\\
    &= \sothreesum{a}\composition{\adgen{a}}{\commutator{\adgen{a}}{\leftgen{b}}}+\composition{\commutator{\adgen{a}}{\leftgen{b}}}{\adgen{a}}\\
    &= \sothreesum{a}\composition{\adgen{a}}{\leftliegen{a}{b}}+\composition{\leftliegen{a}{b}}{\adgen{a}}\\
    &= \sothreesum{a,c}\sothreestructureconstants{a}{b}{c}\big(\composition{\adgen{a}}{\leftgen{c}}+\composition{\leftgen{c}}{\adgen{a}}\big)\\
    &= \sothreesum{a,c}\sothreestructureconstants{a}{b}{c}\Big(\composition{\big(\leftgen{a}-\rightgen{a}\big)}{\leftgen{c}}+\composition{\leftgen{c}}{\big(\leftgen{a}-\rightgen{a}\big)}\Big)\\
    &= \sothreesum{a,c}\sothreestructureconstants{a}{b}{c}\big(\composition{\leftgen{a}}{\leftgen{c}}+\composition{\leftgen{c}}{\leftgen{a}}\big)-2\sothreesum{a,c}\sothreestructureconstants{a}{b}{c}\,\composition{\leftgen{c}}{\rightgen{a}}.
\end{align*}
\noindent Since total contraction of a symmetric and antisymmetric object yields zero we find:
\begin{equation}\label{eqn:first-e-l-commutator}
    \commutator{\casaction}{\leftgen{b}} = \definition{-2\sothreesum{c,a}\sothreestructureconstants{b}{c}{a}\,\composition{\leftgen{c}}{\rightgen{a}}}{-2F(\generator[b])}
\end{equation}
\noindent If we instead had calculated $\commutator{\casaction}{\rightgen{a}}$ we would discover that:
\begin{equation}\label{eqn:l-r-e-commutator}
    \commutator{\casaction}{\leftgen{a}} = \commutator{\casaction}{\rightgen{a}}
\end{equation}

Next let us consider:
\begin{equation*}
    \commutator{\casaction}{F(\generator[b])} = \sothreesum{c,a}\sothreestructureconstants{b}{c}{a}\commutator{\casaction}{\composition{\leftgen{c}}{\rightgen{a}}}.
\end{equation*}
\noindent From \eqref{eqn:left-central} and \eqref{eqn:cas-action-identity} we see:
\begin{align*}
    \commutator{\casaction}{F(\generator[b])} &= -2\sothreesum{d,c,a}\sothreestructureconstants{b}{c}{a}\commutator{\composition{\leftgen{d}}{\rightgen{d}}}{\composition{\leftgen{c}}{\rightgen{a}}}\\
    &= -2\sothreesum{d,c,a}\sothreestructureconstants{b}{c}{a}\big(\composition{\leftgen{d}}{\leftgen{c}}{\commutator{\rightgen{d}}{\rightgen{a}}}+\composition{\commutator{\leftgen{d}}{\leftgen{c}}}{\rightgen{a}}{\rightgen{d}}\big)\\
    &= -2\sothreesum{d,c,a}\sothreestructureconstants{b}{c}{a}\big(\composition{\leftgen{d}}{\leftgen{c}}{\rightliegen{a}{d}}+\composition{\leftliegen{d}{c}}{\rightgen{a}}{\rightgen{d}}\big)\\
    &= -2\sothreesum{e,d,c,a}\sothreestructureconstants{b}{c}{a}\sothreestructureconstants{a}{d}{e}\composition{\leftgen{d}}{\leftgen{c}}{\rightgen{e}}-2\sothreesum{e,d,c,a}\sothreestructureconstants{b}{c}{a}\sothreestructureconstants{d}{c}{e}\composition{\leftgen{e}}{\rightgen{a}}{\rightgen{d}}.
\end{align*}
\noindent Utilising:
\begin{equation}\label{eqn:eps-identity}
    \sothreesum{x}\sothreestructureconstants{x}{p}{q}\sothreestructureconstants{x}{r}{s}=\delta_{pr}\delta_{qs}-\delta_{ps}\delta_{qr}
\end{equation}
\noindent we find:
\begin{align*}
    \commutator{\casaction}{F(\generator[b])} &= -2\sothreesum{e,d,c}(\delta_{bd}\delta_{ce}-\delta_{be}\delta_{cd})\composition{\leftgen{d}}{\leftgen{c}}{\rightgen{e}}\\
    &\quad\quad\quad-2\sothreesum{e,d,a}(\delta_{bd}\delta_{ae}-\delta_{be}\delta_{ad})\composition{\leftgen{e}}{\rightgen{a}}{\rightgen{d}}\\
    &= -2\sothreesum{c}\composition{\leftgen{b}}{\leftgen{c}}{\rightgen{c}} +2\sothreesum{c}\composition{\leftgen{c}}{\leftgen{c}}{\rightgen{b}}\\
    &\quad\quad\quad-2\sothreesum{a}\composition{\leftgen{a}}{\rightgen{a}}{\rightgen{b}}+2\sothreesum{a}\composition{\leftgen{b}}{\rightgen{a}}{\rightgen{a}}\\
    &= \composition{\leftgen{b}}{(\casaction-2\leftcas)}+2\composition{\leftcas}{\rightgen{b}}\\
    &\quad\quad\quad+\composition{(\casaction-2\leftcas)}{\rightgen{b}}+2\composition{\leftgen{b}}{\leftcas}\\
\end{align*}
\noindent and thus:
\begin{equation}\label{eqn:e-f-commutator}
    \commutator{\casaction}{F(\generator[b])} = \composition{\leftgen{b}}{\casaction}+
    \composition{\casaction}{\rightgen{b}} = \composition{\rightgen{b}}{\casaction}+
    \composition{\casaction}{\leftgen{b}}
\end{equation}
\noindent with the final equality following from \eqref{eqn:l-r-e-commutator}.

Hence, combining \eqref{eqn:first-e-l-commutator} and \eqref{eqn:e-f-commutator}:
\begin{align*}
    \commutator{\casaction}{\commutator{\casaction}{\commutator{\casaction}{\leftgen{b}}}} &= -2\commutator{\casaction}{\commutator{\casaction}{F(\generator[b])}}\\
    &= -2\commutator{\casaction}{\composition{\big(\leftgen{b}}{\casaction}+
    \composition{\casaction}{\rightgen{b}}\big)}\\
    &= -2\composition{\commutator{\casaction}{\leftgen{b}}}{\casaction}-2
    \composition{\casaction}{\commutator{\casaction}{\rightgen{b}}}\\
    &= -2\big(\composition{\commutator{\casaction}{\leftgen{b}}}{\casaction}+\composition{\casaction}{\commutator{\casaction}{\leftgen{b}}}\big)
\end{align*}
\noindent thus, finally:
\begin{equation}
    \commutator{\casaction}{\commutator{\casaction}{\commutator{\casaction}{\leftgen{b}}}} = -2\commutator{\casaction^2}{\leftgen{b}}\quad_{\blacksquare}
\end{equation}

\section{Derivation of the Images of Multipoles under Step-Level and Step-Down}\label{app:multipole-step-image-proof}

Here we will prove the results given in \eqref{eqn:step-down-multipole-image} and \eqref{eqn:step-level-multipole-image}.

\subsection{Step-Level Image}

The step-level by $v\in\sothree$ of a multipole $\multipole{k}$ is given by:
\begin{equation*}
    \composition{\steplevel{v}}{\multipole{k}}=\composition{\frac{\composition{\casfactoraction{k-1}}{\casfactoraction{k+1}}}{-4k(k+1)}}{\sothreeleft[v]}{\multipole{k}}.
\end{equation*}
\noindent Commuting through the $\casfactoraction{\cdot}$ we find:
\begin{equation*}
    ... = \frac{1}{-4k(k+1)}\composition{\Big(\commutator{E}{\commutator{E}{\sothreeleft[v]}}+2\commutator{E}{\sothreeleft[v]}-4k(k+1)\sothreeleft[v]\Big)}{\multipole{k}}.
\end{equation*}
\noindent From \eqref{eqn:e-f-commutator} we see that:
\begin{equation}\label{eqn:e-e-l-identity}
\begin{aligned}
    \commutator{E}{\commutator{E}{\sothreeleft[v]}}&=-2\big(\composition{\sothreeright[v]}{\casaction}+
    \composition{\casaction}{\sothreeleft[v]}\big)\\
    &=-2\commutator{E}{\sothreeleft[v]}-2\big(\composition{\sothreeright[v]}{\casaction}+
    \composition{\sothreeleft[v]}{\casaction}\big)
\end{aligned}
\end{equation}
\noindent which we combine with the previous equation and \eqref{eqn:adjoint-l-r} to find:
\begin{equation}\label{eqn:left-step-level-ad}
\begin{aligned}
    \composition{\steplevel{v}}{\multipole{k}} &= \frac{1}{-4k(k+1)}\composition{\Big(-2\composition{\big(\sothreeright[v]+
    \sothreeleft[v]\big)}{\casaction}-4k(k+1)\sothreeleft[v]\Big)}{\multipole{k}}\\
    &= \frac{-2k(k+1)}{-4k(k+1)}\composition{\big(
    \sothreeleft[v]-\sothreeright[v]\big)}{\multipole{k}}\\
    &= \half\composition{\sothreead[v]}{\multipole{k}}\\
    &= \half\composition{\multipole{k}}{\sothreead[v]}
\end{aligned}
\end{equation}
\noindent from which \eqref{eqn:step-level-multipole-image} follows.

\subsection{Step-Down Image}

The step-down by $\generator[a]\in\sothree$ of a multipole $\multipole{k}$ is given by:
\begin{align*}
    \composition{\stepdown{\generator[a]}}{\multipole{k}}&=\composition{\frac{\composition{\casfactoraction{k}}{\casfactoraction{k+1}}}{4k(2k+1)}}{\leftgen{a}}{\multipole{k}}\\
    &= \composition{\frac{\big(\casaction^2+2(k+1)^2E+k(k+1)^2(k+2)\id\big)}{4k(2k+1)}}{\leftgen{a}}{\multipole{k}}
\end{align*}
\noindent To proceed, let us consider how $\casaction$ and $\casaction^2$ act on $\composition{\leftgen{a}}{\multipole{k}}$.

First, we see from \eqref{eqn:adjoint-l-commutation}:
\begin{align*}
    \composition{\adgen{c}}{\leftgen{a}}{\multipole{k}\Big(\bigotimes_{j=1}^{k}\generator[b_j]\Big)} &= \composition{\sothreeleft[\adgen{c}(\generator[a])]}{\multipole{k}\Big(\bigotimes_{j=1}^{k}\generator[b_j]\Big)} \\ 
    &\phantom{=}+ \composition{\leftgen{a}}{\adgen{c}}{\multipole{k}\Big(\bigotimes_{j=1}^{k}\generator[b_j]\Big)}
\end{align*}
\noindent and so:
\begin{align*}
    \composition{\casaction&}{\leftgen{a}}{\multipole{k}\Big(\bigotimes_{j=1}^{k}\generator[b_j]\Big)}\\
    &= \sothreesum{c}\composition{\adgen{c}}{\adgen{c}}{\leftgen{a}}{\multipole{k}\Big(\bigotimes_{j=1}^{k}\generator[b_j]\Big)}\\
    &= (-2-k(k+1))\composition{\leftgen{a}}{\multipole{k}\Big(\bigotimes_{j=1}^{k}\generator[b_j]\Big)} + 2\sothreesum{c}\composition{\sothreeleft[\adgen{c}(\generator[a])]}{\adgen{c}}{\multipole{k}\Big(\bigotimes_{j=1}^{k}\generator[b_j]\Big)}.
\end{align*}
\noindent We may evaluate this second term with the help of \eqref{eqn:eps-identity}:
\begin{align*}
     2\sothreesum{c}&\composition{\sothreeleft[\adgen{c}(\generator[a])]}{\adgen{c}}{\multipole{k}\Big(\bigotimes_{j=1}^{k}\generator[b_j]\Big)}\\
     &= 2\sum_{p=1}^{k}\sothreesum*[r]{c,d,e}\sothreestructureconstants{c}{a}{d}\sothreestructureconstants{c}{b_{p}}{e}\composition{\leftgen{d}}{\multipole{k}\Big(\tensor{\generator[e];\smashoperator{\bigotimes_{j=1,j\neq p}^{k}}\generator[b_j]}\Big)}\\
     &= 2\sum_{p=1}^{k}\sothreesum*[r]{d}
     \delta_{ab_{p}}\composition{\leftgen{d}}{\multipole{k}\Big(\tensor{\generator[d];\smashoperator{\bigotimes_{j=1,j\neq p}^{k}}\generator[b_j]}\Big)} - 2\sum_{p=1}^{k}\composition{\leftgen{b_{p}}}{\multipole{k}\Big(\tensor{\generator[a];\smashoperator{\bigotimes_{j=1,j\neq p}^{k}}\generator[b_j]}\Big)}.
\end{align*}
\noindent For notational convenience let us denote:
\begin{align}
    \definition{A&}{\composition{\leftgen{a}}{\multipole{k}\Big(\bigotimes_{j=1}^{k}\generator[b_j]\Big)}}\\
    \definition{\composition{B}{\leftgen{a}}{\multipole{k}\Big(\bigotimes_{j=1}^{k}\generator[b_j]\Big)}&}{\sum_{p=1}^{k}\sothreesum*[r]{d}
    \delta_{ab_{p}}\composition{\leftgen{d}}{\multipole{k}\Big(\tensor{\generator[d];\smashoperator{\bigotimes_{j=1,j\neq p}^{k}}\generator[b_j]}\Big)}}\label{eqn:b-operator}\\
    \definition{\composition{C}{\leftgen{a}}{\multipole{k}\Big(\bigotimes_{j=1}^{k}\generator[b_j]\Big)}&}{\sum_{p=1}^{k}\composition{\leftgen{b_{p}}}{\multipole{k}\Big(\tensor{\generator[a];\smashoperator{\bigotimes_{j=1,j\neq p}^{k}}\generator[b_j]}\Big)}}
\end{align}
\noindent so we may write:
\begin{equation}
    \composition{\casaction}{A}=\composition{\Big(\big(-2-k(k+1)\big)\id+2B-2C\Big)}{A}
\end{equation}

Applying a second $\casaction$, we may reuse the results so far to find:
\begin{align*}
    \composition{\casaction}{B}{A}&=-k(k-1)\composition{B}{A}\\
    \composition{\casaction}{C}{A}&=\composition{\big(-2k\id+2B-k(k+3)C+2D\big)}{A}
\end{align*}
\noindent where we have defined:
\begin{equation}\label{eqn:d-operator}
\begin{aligned}
    \definition{\composition{D}{\leftgen{a}&}{\multipole{k}\Big(\bigotimes_{j=1}^{k}\generator[b_j]\Big)}\\}{&\begin{cases}
        \displaystyle{\sum_{p=1}^{k}\sum_{q=1,q\neq p}^{k}\sothreesum*[r]{d}\delta_{b_{p}b_{q}}\composition{\leftgen{d}}{\multipole{k}\Big(\tensor{\generator[a];\generator[d];\smashoperator{\bigotimes_{j=1,j\neq p,q}^{k}}\generator[b_j]}\Big)}} & k>1\\
        0 & k=1.
    \end{cases}}
\end{aligned}
\end{equation}
\noindent Thus we find:
\begin{equation}
    \composition{E}{E}{A}=\composition{\big((k+1)^2(k^2+4)\id-4(k^2+2)B+4(k+1)^2C-4D\big)}{A}
\end{equation}
\noindent and so:
\begin{equation}\label{eqn:step-down-on-multipole}
    \composition{\stepdown{\generator[a]}}{\multipole{k}\Big(\bigotimes_{j=1}^{k}\generator[b_j]\Big)}=\composition{\frac{\big((2k-1)B-D\big)}{k(2k+1)}}{A}.
\end{equation}
\noindent To progress further we must extract the summed $\generator[d]$ from within $\multipole{k}$ in \eqref{eqn:b-operator} and \eqref{eqn:d-operator}. We do this by using the multipole recursion relation \eqref{eqn:recursivestepmultipole}, and an identity we shall now derive.

For $k=1$, we see:
\begin{equation*}
    \sothreesum{e}\composition{\leftgen{e}}{\multipole{1}(\generator[e])}=\sothreesum{e}\composition{\leftgen{e}}{\leftgen{e}}{\multipole{0}(1)}=\composition{\leftcas}{\multipole{0}(1)}.
\end{equation*}
\noindent For $k>1$, consider:
\begin{align*}
    \sothreesum{e}\composition{\leftgen{e}&}{\multipole{k}\Big(\tensor{\generator[e];\bigotimes_{j=1}^{k-1}\generator[c_j]}\Big)}\\
    &=\frac{1}{4k(2k-1)}\sothreesum{e}\composition{\leftgen{e}}{\casfactoraction{k-2}}{\casfactoraction{k-1}}{\leftgen{e}}{\multipole{k-1}\Big(\bigotimes_{j=1}^{k-1}\generator[c_j]\Big)}\\
    &=\frac{1}{4k(2k-1)}\sothreesum{e}\composition{\leftgen{e}}{\casfactoraction{k-2}}{\commutator{\casaction}{\leftgen{e}}}{\multipole{k-1}\Big(\bigotimes_{j=1}^{k-1}\generator[c_j]\Big)}+0\\
    &=\frac{1}{4k(2k-1)}\sothreesum{e}\composition{\leftgen{e}}{\Big(\commutator{\casaction}{\commutator{\casaction}{\leftgen{e}}}-2(k-1)\commutator{\casaction}{\leftgen{e}}\Big)}{\multipole{k-1}\Big(\bigotimes_{j=1}^{k-1}\generator[c_j]\Big)}
\end{align*}
\noindent and by \eqref{eqn:e-e-l-identity}, \eqref{eqn:first-e-l-commutator} and \eqref{eqn:cas-action-identity} we find:
\begin{align*}
    ...&=\frac{1}{4k(2k-1)}\sothreesum{e}\composition{\leftgen{e}}{\Big(-2\composition{\big(\leftgen{e}+\rightgen{e}\big)}{\casaction}-2k\commutator{\casaction}{\leftgen{e}}\Big)}{\multipole{k-1}\Big(\bigotimes_{j=1}^{k-1}\generator[c_j]\Big)}\\
    ...&=\frac{1}{4k(2k-1)}\composition{\Big(4k(k-1)\leftcas-k(k-1)\casaction+4k\sothreesum*{e,g,a}\sothreestructureconstants{e}{g}{a}\composition{\leftgen{e}}{\leftgen{g}}{\rightgen{a}}\Big)}{\multipole{k-1}\Big(\bigotimes_{j=1}^{k-1}\generator[c_j]\Big)}\\
    ...&=\frac{1}{4k(2k-1)}\composition{\Big(4k(k-1)\leftcas+k^2(k-1)^2\id+4k\sothreesum{e}\composition{\leftgen{e}}{\rightgen{e}}\Big)}{\multipole{k-1}\Big(\bigotimes_{j=1}^{k-1}\generator[c_j]\Big)}\\
    ...&=\frac{1}{4k(2k-1)}\composition{\Big(4k(k-1)\leftcas+k^2(k-1)^2\id-2k\casaction+4k\leftcas\Big)}{\multipole{k-1}\Big(\bigotimes_{j=1}^{k-1}\generator[c_j]\Big)}\\
    ...&=\frac{k}{4(2k-1)}\composition{L\big(4\sothreecasimirelement+(k-1)(k+1)\big)}{\multipole{k-1}\Big(\bigotimes_{j=1}^{k-1}\generator[c_j]\Big)}.
\end{align*}
\noindent This result is consistent with the $k=1$ case. Applying this to $B$ and $D$ in \eqref{eqn:step-down-on-multipole} we find the identity \eqref{eqn:step-down-multipole-image}.

\subsection{Right Multiplication Images}

The results of the previous subsections may be utilised to derive the form of a right multiplication of a multipole. This is essential to expand the multiplication table of Appendix \ref{app:multipole-multiplication}.

We observe that from the definition of $\sothreead[v]$ where $v\in\sothree$:
\begin{align*}
    0=\composition{\sothreead[v]}{\casfactoraction{k}}{\multipole{{k}}}=\composition{\casfactoraction{k}}{\sothreead[v]}{\multipole{{k}}}=\composition{\casfactoraction{k}}{\big(\sothreeleft[v]-\sothreeright[v]\big)}{\multipole{{k}}}
\end{align*}
\noindent and so:
\begin{equation*}
    \composition{\casfactoraction{k}}{\sothreeleft[v]}{\multipole{{k}}}=\composition{\casfactoraction{k}}{\sothreeright[v]}{\multipole{{k}}}
\end{equation*}
\noindent which implies that:
\begin{align}
    \composition{\stepdownright{v}}{\multipole{k}}&=\composition{\stepdown{v}}{\multipole{k}}\\
    \composition{\stepupright{v}}{\multipole{k}}&=\composition{\stepup{v}}{\multipole{k}}.
\end{align}
\noindent While:
\begin{align*}
    \composition{\steplevelright{v}}{\multipole{k}}&=\composition{\frac{\composition{\casfactoraction{k-1}}{\casfactoraction{k+1}}}{-4k(k+1)}}{\sothreeright[v]}{\multipole{k}}\\
    &=\composition{\frac{\composition{\casfactoraction{k-1}}{\casfactoraction{k+1}}}{-4k(k+1)}}{\big(\sothreeleft[v]-\sothreead[v]\big)}{\multipole{k}}\\
    &=\composition{\steplevel{v}}{\multipole{k}}-\composition{\sothreead[v]}{\multipole{k}}\\
    &=-\half\composition{\sothreead[v]}{\multipole{k}}
\end{align*}
\noindent which from \eqref{eqn:left-step-level-ad} gives:
\begin{equation}
    \composition{\steplevelright{v}}{\multipole{k}}=-\composition{\steplevel{v}}{\multipole{k}}.
\end{equation}

\section{Proof of Scalar Multiple Casimir Action}\label{app:proof-scalar-casimir-action}

Here we will prove that on the quotient algebra $\spinalgebra{\half[k]}$:
\begin{equation*}
    \definition{\spinalgebra{\half[k]}}{\frac{\usothree}{\ideal{\image{\multipole{k+1}}}}}
\end{equation*}
\noindent of $\usothree$ by the ideal generated by $\image{\multipole{k+1}}$, $k\in\naturals$, that the Casimir element $\sothreecasimirelement$ acts as a scalar:
\begin{equation*}
    \leftcas=L\left(\frac{k(k+2)}{4}\right)
\end{equation*}

To do this, consider $\image{f}$ where:
\begin{equation}\label{eqn:stepcombination}
    \definition{f}{\mapdefinition{\tensor{\generator[a];\bigotimes_{j=1}^{k+1}\generator[b_{j}]}}{\composition{\stepdown{\generator[a]}}{\stepup{\generator[b_{1}]}}{\multipole{k}\Big(\bigotimes_{j=2}^{k+1}\generator[b_{j}]\Big)}}}
\end{equation}
\noindent From \eqref{eqn:recursivestepmultipole} we know:
\begin{equation}\label{eqn:step-combination-zero}
    \composition{\composition{\stepup{\generator[b_{1}]}}{\multipole{k}(A)}}=\multipole{k+1}(\tensor{\generator[b_{1}];A})=0
\end{equation} 
\noindent with the final equality following from $\multipole{k+1}=0$ in $\spinalgebra{\half[k]}$, and thus $f=0$. However, we know from the main analysis of $\usothree$ that $\image{f}$ can be written as a linear combination of central multiples of $\multipole{k}$. Since $\multipole{k}$ is non-zero in $\spinalgebra{\half[k]}$, and $\image{f}$ is non-trivial in $\usothree$, there must be some new identity amongst the central multiples causing $\image{f}$ to be trivial.

Equations \eqref{eqn:stepcombination} and \eqref{eqn:step-combination-zero} show that we are studying a step-down from the multipole $\multipole{k+1}$. When $k=0$:
\begin{equation}
    0=f(\tensor{\generator[a];\generator[b]})=\third\sothreecasimirelement\delta_{ab}=\third\composition{\leftcas}{\multipole{0}}(1)
\end{equation}
\noindent from \eqref{eqn:step-down-dipole-image}. Since $\spinalgebra{0}$ is spanned by $\multipole{0}$ we conclude that on the whole of $\spinalgebra{0}$:
\begin{equation}
    \leftcas=0.
\end{equation}
\noindent If $k>0$ we use \eqref{eqn:step-down-multipole-image} to find:
\begin{equation*}
\begin{aligned}
    0&=\composition{\frac{L\big(4\sothreecasimirelement+k(k+2)\big)}{4(2k+3)(2k+1)}}{\sum_{p=1}^{k+1}\Bigg((2k+1)\delta_{ab_{p}}\multipole{k}\Big(\bigotimes_{j\neq p}\generator[b_{j}]\Big)
    -\smashoperator{\sum_{q=1,q\neq p}^{k+1}}\delta_{b_{p}b_{q}}\multipole{k}\Big(\tensor{\generator[a];\smashoperator{\bigotimes_{j\neq p,q}}\generator[b_{j}]}\Big)\Bigg)}\\
    &=\composition{\frac{L\big(4\sothreecasimirelement+k(k+2)\big)}{4(2k+3)(2k+1)}}{\multipole{k}\Bigg(\sum_{p=1}^{k+1}\smashoperator[r]{\sum_{q=1,q\neq p}^{k+1}}\tensor{\Big(\frac{2k+1}{k}\delta_{ab_{p}}\generator[b_{q}]
    -\delta_{b_{p}b_{q}}\generator[a]\Big);\smashoperator{\bigotimes_{j\neq p,q}}\generator[b_{j}]}\Bigg)}.
\end{aligned}
\end{equation*}
\noindent In $\usothree$ the prefactor in the above has trivial kernel, since $\usothree$ contains no zero divisors\cite{dixmier_enveloping_1977}. Thus, if an element of the above is non-zero in $\usothree$ it is because $\multipole{k}$ is non-zero on its argument. Furthermore, there are necessarily enough arguments $\set{C_{j}}$ such that $\set{\multipole{k}(C_{j})}$ span $\image{f}$, and therefore $\image{\multipole{k}}$. Since $\image{\multipole{k}}$ is non-trivial in $\spinalgebra{\half[k]}$, and $k>0$ we must have that:
\begin{equation}\label{eqn:casimir-scalar-top-multipole}
    \composition{L\big(4\sothreecasimirelement+k(k+2)\big)}{\multipole{k}}=0
\end{equation}

As $4\sothreecasimirelement+k(k+2)$ is central we may repeat the process by stepping-down \eqref{eqn:casimir-scalar-top-multipole}, producing a family of identities $\forall n\in\set{0,...,k}$:
\begin{equation}\label{eqn:casimir-scalar-identity-family}
    \composition{\left(\operatorname*{\mathlarger{\mathlarger{\bigcirc}}}_{j=n}^{k}L\big(4\sothreecasimirelement+j(j+2)\big)\right)}{\multipole{n}}=0
\end{equation}
\noindent where $\operatorname*{\mathlarger{\mathlarger{\bigcirc}}}_{j=n}^{k}$ denotes composition over the indexed maps.

A priori, any combination of these $L\big(4\sothreecasimirelement+j(j+2)\big)$ could be responsible for annihilating $\image{\multipole{n}}$. To make progress, let us first consider a non-empty subset $I\subset\set{0,...,k-1}$ and suppose:
\begin{equation}\label{eqn:no-k-annihilation-hypothesis}
    \composition{\left(\operatorname*{\mathlarger{\mathlarger{\bigcirc}}}_{j\in I}L\big(4\sothreecasimirelement+j(j+2)\big)\right)}{\multipole{n}}=0
\end{equation}
\noindent for some $n\in\set{0,...,k}$. Since $\multipole{k}$ may be written as a series of step-ups from $\multipole{n}$ by \eqref{eqn:recursivestepmultipole}, we may use the fact that the composition in \eqref{eqn:no-k-annihilation-hypothesis} commutes with step-ups to find:
\begin{equation}\label{eqn:no-k-annihilation-hypothesis-on-k-multipole}
    \composition{\left(\operatorname*{\mathlarger{\mathlarger{\bigcirc}}}_{j\in I}L\big(4\sothreecasimirelement+j(j+2)\big)\right)}{\multipole{k}}=0
\end{equation}
\noindent Now, observe that for any $p$ we may write:
\begin{equation}\label{eqn:k-in-terms-of-p-cas}
    L\big(4\sothreecasimirelement+k(k+2)\big) = L\big(4\sothreecasimirelement+p(p+2)\big) + L\big((k-p)(k+p+2)\big),
\end{equation}
\noindent which we may use to rewrite \eqref{eqn:no-k-annihilation-hypothesis-on-k-multipole} as:
\begin{equation}\label{eqn:all-k-composition-on-k-multipole}
    \composition{\left(\operatorname*{\mathlarger{\mathlarger{\bigcirc}}}_{j\in I}\big(L\big(4\sothreecasimirelement+k(k+2)\big) - L\big((k-j)(k+j+2)\big)\big)\right)}{\multipole{k}}=0.
\end{equation}
\noindent We note that since $k\notin I$ that there is a left multiplication of a non-zero scalar in \eqref{eqn:all-k-composition-on-k-multipole}. Thus, from \eqref{eqn:casimir-scalar-top-multipole} we find:
\begin{equation}
    \Big(\prod_{j\in I}-(k-j)(k+j+2)\Big)\multipole{k}=0,
\end{equation}
\noindent which implies $\image{\multipole{k}}$ is trivial. This is in contradiction with our construction of $\spinalgebra{\half[k]}$ and thus \eqref{eqn:no-k-annihilation-hypothesis} must be impossible $\forall n\in\set{0,...,k}$. This means any annihilating action of a composition of factors $L\big(4\sothreecasimirelement+p(p+2)\big)$ must include the factor with $p=k$.

With this in hand, let us consider the identity \eqref{eqn:casimir-scalar-identity-family} on $\multipole{0}$:
\begin{equation}\label{eqn:casimir-scalar-identity-monopole}
    \composition{\left(\operatorname*{\mathlarger{\mathlarger{\bigcirc}}}_{j=0}^{k}L\big(4\sothreecasimirelement+j(j+2)\big)\right)}{\multipole{0}}=0
\end{equation}
\noindent and notice that it is an annihilating polynomial for $\leftcas$ on $\image{\multipole{0}}$. Thus, using the results of section \ref{sec:real-operator-formalism} we may resolve the identity on $\image{\multipole{0}}$:
\begin{equation}
    \multipole{0}=\sum_{j=0}^{k}\left(\operatorname*{\mathlarger{\mathlarger{\bigcirc}}}_{p=0,p\neq j}^{k}\composition{\frac{L\big(4\sothreecasimirelement+p(p+2)\big)}{-j(j+2)+p(p+2)}\right)}{\multipole{0}}=\sum_{j=0}^{k}\Pi_{j}.
\end{equation}
\noindent From our earlier argument, no annihilating composition like \eqref{eqn:no-k-annihilation-hypothesis} can exist, and thus we must conclude that $\image{\multipole{0}}\cap\image{\Pi_{k}}\neq\set{0}$, since $\Pi_{k}$ contains no factor $L\big(4\sothreecasimirelement+k(k+2)\big)$ by definition. However, $\dim(\image{\multipole{0}})=1$, and since $\Pi_{k}$ is linear, we must conclude that $\image{\multipole{0}}\subset\image{\Pi_{k}}$ and $\image{\multipole{0}}\cap\image{\Pi_{j}}=\set{0}$ for $j\neq k$ by orthogonality.

Thus for $j\neq k$, $\Pi_{j}=0$, and so we have a family of annihilating polynomials for $\leftcas$ on $\image{\multipole{0}}$, $\forall j\in\set{0,...,k-1}$:
\begin{equation}\label{eqn:monopole-annihilating-family}
    \left(\operatorname*{\mathlarger{\mathlarger{\bigcirc}}}_{p=0,p\neq j}^{k}\composition{L\big(4\sothreecasimirelement+p(p+2)\big)\right)}{\multipole{0}}=0.
\end{equation}
\noindent But every annihilating polynomial must be a polynomial multiple of the minimal polynomial\cite{axler_linear_2014}. Since the family \eqref{eqn:monopole-annihilating-family} have only one factor in common, we must conclude that:
\begin{equation}\label{eqn:casimir-scalar-monopole}
    \composition{L\big(4\sothreecasimirelement+k(k+2)\big)}{\multipole{0}}=0.
\end{equation}

By the recursive relationship between the multipoles \eqref{eqn:recursivestepmultipole}, all $\multipole{k}$ begin from repeated stepping-up from $\multipole{0}$. $L\big(4\sothreecasimirelement+k(k+2)\big)$ is commutative with step-ups, thus we find $\forall n\in\set{0,...,k}$:
\begin{equation}\label{eqn:casimir-scalar-all-multipoles}
    \composition{L\big(4\sothreecasimirelement+k(k+2)\big)}{\multipole{n}}=0.
\end{equation}
\noindent Since the multipoles $\set{\multipole{n}|\,n\in\set{0,...,k}}$ form a basis for $\spinalgebra{\half[k]}$, from \eqref{eqn:casimir-scalar-all-multipoles} we must finally conclude that on the whole of $\spinalgebra{\half[k]}$:
\begin{equation}
    L\big(4\sothreecasimirelement+k(k+2)\big)=0\quad_{\blacksquare}
\end{equation}

\end{document}